\documentclass[11pt,a4paper]{article}
 \usepackage{jheppub}

\pdfoutput=1
\usepackage{amsfonts}
\usepackage{amssymb}
\usepackage{amsmath}
\usepackage{graphicx}
\usepackage{booktabs}
\usepackage{array}
\usepackage{color}
\usepackage{ulem}
\usepackage[small,bf]{caption}
\usepackage{mciteplus}
\usepackage[usenames,dvipsnames]{xcolor}
\usepackage{subfigure}
\usepackage{verbatim}
\usepackage{dsfont}
\usepackage{slashed}
\usepackage{color,url}

\def\1{\mathbf{1}}
\def\3{\mathbf{3}}
\def\2{\mathbf{2}}  
\def\a{\alpha}

\def\g{\gamma}
\def\G{\Gamma}

\def\D{\Delta}
\def\e{\varepsilon}
\def\k{\kappa}

\def\y{\eta}
\def\o{\omega}
\def\O{\Omega}
\def\x{\phi}
\def\f{\phi}

\newcommand{\keV}{{\rm keV}}

\newcommand{\GeV}{{\rm GeV}}
\newcommand{\TeV}{{\rm TeV}}
\newcommand{\vev}[1]{\langle #1 \rangle}

\allowdisplaybreaks[1]

\numberwithin{equation}{section}

\begin{document}

\thispagestyle{empty}
\begin{flushright}
FERMILAB-PUB-17-285-T\,\,\,\, \\
FTUAM-17-14\,\,\,\,\\ IFT-UAM/CSIC-17-070\\

\vspace*{2.mm} 
\end{flushright}

\title{\color{blue} Dark Matter and the elusive $\mathbf{Z'}$\\ in a dynamical Inverse Seesaw scenario}

\author[a]{Valentina~De~Romeri,}
\author[b ,c]{Enrique~Fernandez-Martinez,}
\author[b, c]{Julia~Gehrlein,}
\author[d]{Pedro~A.~N.~Machado}
\author[e]{and Viviana~Niro}

\affiliation[a]{
      AHEP Group, Instituto de F\'{\i}sica Corpuscular,
      C.S.I.C./Universitat de Val\`encia,  \\
      Calle Catedr\'atico Jos\'e Beltr\'an, 2 E-46980 Paterna, Spain}
   \affiliation[b]{   Departamento  de  F\'{\i}sica Te\'{o}rica,  Universidad  Aut\'{o}noma  de  Madrid,\\
\it Cantoblanco  E-28049  Madrid,  Spain}
\affiliation[c]{ Instituto  de  F\'{\i}sica  Te\'{o}rica  UAM/CSIC,\\
 Calle Nicol\'{a}s Cabrera  13-15,  Cantoblanco  E-28049  Madrid,  Spain}
\affiliation[d]{ Theoretical Physics Department, Fermi National Accelerator Laboratory, Batavia, IL, 60510, USA }
\affiliation[e]{Institut f\"{u}r Theoretische Physik, Ruprecht-Karls-Universit\"{a}t Heidelberg,
Philosophenweg 16, 69120 Heidelberg, Germany}

\emailAdd{ deromeri@ific.uv.es, enrique.fernandez-martinez@uam.es, julia.gehrlein@uam.es, pmachado@fnal.gov, 
niro@thphys.uni-heidelberg.de}

\abstract{The Inverse Seesaw naturally explains the smallness of neutrino masses via an approximate $B-L$ symmetry broken only by a correspondingly small parameter. In this work the possible dynamical generation of the Inverse Seesaw neutrino mass mechanism from the spontaneous breaking of a gauged $U(1)$ $B-L$ symmetry is investigated. Interestingly, the Inverse Seesaw pattern requires a chiral content such that anomaly cancellation predicts the existence of extra fermions belonging to a dark sector with large, non-trivial, charges under the $U(1)$ $B-L$. We investigate the phenomenology associated to these new states and find that one of them is a 
viable dark matter candidate with mass around the TeV scale, whose interaction with the Standard Model is mediated by the $Z'$ boson associated to the gauged $U(1)$ $B-L$ symmetry. 
Given the large charges required for anomaly cancellation in the dark sector, the $B-L$ $Z'$ interacts preferentially with this dark sector rather than with the Standard Model. 
This suppresses the rate at direct detection searches and thus alleviates the constraints on $Z'$-mediated dark matter relic abundance. The collider phenomenology of this \textit{elusive} $Z'$ is also discussed.}


\maketitle

\section{Introduction}

The simplest and most popular mechanism to accommodate the evidence for neutrino masses and 
mixings~\cite{Tortola:2013voa,Forero:2014bxa,Gonzalez-Garcia:2014bfa,Gonzalez-Garcia:2015qrr,Capozzi:2016rtj,Esteban:2016qun} 
and to naturally explain their extreme smallness, calls upon the introduction of right-handed neutrinos through the celebrated 
Seesaw mechanism~\cite{Minkowski:1977sc,Ramond:1979py,GellMann:1980vs,Yanagida:1979as,Mohapatra:1979ia,Schechter:1980gr}. Its appeal stems from the simplicity of its particle content, consisting only of the right-handed neutrinos otherwise conspicuously missing from the Standard Model (SM) ingredients. 
In the Seesaw mechanism, the smallness of neutrino masses is explained through the ratio of their Dirac masses and the Majorana mass term of the extra fermion singlets. Unfortunately, this very same ratio suppresses any phenomenological probe of the existence of this mechanism. Indeed, either the right-handed neutrino masses would be too large to be reached by our highest energy colliders, or the Dirac masses, and hence the Yukawa interactions that mediate the right-handed neutrino phenomenology, would be too small for even our more accurate precision probes through flavour and precision electroweak observables. 

However, a large hierarchy of scales is not the only possibility to naturally explain the smallness of neutrino masses. Indeed, neutrino masses are protected by the 
$B-L$ (Baryon minus Lepton number) global symmetry, otherwise exact in the SM. Thus, if this symmetry is only mildly broken, neutrino masses will be necessarily suppressed by 
the small $B-L$-breaking parameters. Conversely, the production and detection of the extra right-handed neutrinos at colliders as well as their indirect effects in flavour and precision electroweak observables are not protected by the $B-L$ symmetry and therefore not necessarily suppressed, leading to a much richer and interesting phenomenology. This is the rationale behind the popular Inverse Seesaw Mechanism~\cite{Mohapatra:1986bd} (ISS) as well as the Linear~\cite{Akhmedov:1995vm,Malinsky:2005bi} and Double Seesaw~\cite{Mohapatra:1986bd,Mohapatra:1986aw,Roncadelli:1983ty,Roy:1983be} variants.

In the presence of right-handed neutrinos, $B-L$ is the only flavour-universal SM quantum number that is not anomalous, besides hypercharge. Therefore, just like the addition of right-handed neutrinos, a very natural plausible SM extension is the gauging of this symmetry. In this work these two elements are combined to explore a possible dynamical origin of the ISS pattern from the spontaneous breaking of the gauged $B-L$ symmetry. 

Previous models in the literature have been constructed using the ISS idea or gauging $B-L$ to explain the smallness of the neutrino masses, 
see e.g.~\cite{Klasen:2016qux,Wang:2015saa,Okada:2016gsh,Okada:2016tci,Bandyopadhyay:2017bgh,Cai:2014hka}. 
A minimal model in which the ISS is realised dynamically and where the smallness of the Lepton Number Violating (LNV) term is generated at the two-loop level was studied in~\cite{Bazzocchi:2010dt}. Concerning $U(1)_{B-L}$ extensions of the SM with an ISS  generation of neutrino masses, several models have been investigated~\cite{Khalil:2010iu,Basso:2012ti,Ma:2014qra,Ma:2015raa}. A common origin of both sterile neutrinos and  Dark Matter (DM) has been proposed in~\cite{Escudero:2016tzx,Escudero:2016ksa}. 
An ISS model which incorporates a keV sterile neutrino as a DM candidate was constructed in e.g.~\cite{Abada:2014zra}. 
Neutrino masses break $B-L$, if this symmetry is not gauged and dynamically broken, a massless Goldstone boson, the Majoron, appears in the spectrum. Such models have been investigated 
for example in~\cite{Escudero:2016tzx,Rojas:2017sih}.

Interestingly, since the ISS mechanism requires a chiral pattern in the neutrino sector, the gauging of $B-L$  predicts the existence of extra fermion singlets with non-trivial charges so as to cancel the anomalies. We find that these extra states may play the role of DM candidates as thermally produced 
Weakly Interacting Massive Particles (WIMPs) (see for instance~\cite{Bertone:2004pz,Bertone:2010zza} for a review).

Indeed, the extra states would form a \textit{dark sector}, only connected to the SM via the $Z'$ gauge boson associated to the $B-L$ symmetry and, more indirectly, through the mixing of the scalar responsible for the spontaneous symmetry breaking of $B-L$ with the Higgs boson. For the simplest charge assignment, this dark sector would be constituted by one heavy Dirac and one massless Weyl fermion with large $B-L$ charges. These large charges make the $Z'$ couple preferentially to the dark sector rather than to the SM, making it particularly \textit{elusive}. 
In this work the phenomenology associated with this dark sector and the elusive $Z'$ is investigated. We find that the heavy Dirac fermion of the dark sector can be a viable DM candidate with its relic abundance mediated by the elusive $Z'$. Conversely, the massless Weyl fermion can be probed through measurements of the relativistic degrees of freedom in the early Universe. The collider phenomenology of the elusive $Z'$ is also investigated and 
the LHC bounds are derived. 
  
The paper is structured as follows. In Sec.~\ref{sec:model} we describe the features of the model, namely its Lagrangian and particle content. In Sec.~\ref{sec:DM} we analyse the phenomenology of the DM candidate and its viability. The collider phenomenology of the $Z'$ boson is discussed in Sec.~\ref{sec:colliders}. Finally, in Secs.~\ref{sec:results} and \ref{sec:conclusions} we summarise our results and conclude.

\section{The model}
\label{sec:model}
The usual ISS model consists of the addition of a pair of right-handed SM singlet fermions (right-handed neutrinos) for each massive 
active neutrino~\cite{Mohapatra:1986bd,Wyler:1982dd, Valle:1982yw, Valle:1983dk}. These extra fermion copies, say $N_R$ and $N_R'$, carry a global 
Lepton Number (LN) of $+1$ and $-1$, respectively, and this leads to the following mass Lagrangian 
\begin{equation}
- \mathcal{L}_{\rm ISS} =  \bar L Y_\nu \widetilde{H} N_R + \overline{N_R^c} M_N N_R' + \overline{N_R'^c} \mu\, N_R' + {\rm h.c.},
\end{equation}
where $Y_\nu$ is the neutrino Yukawa coupling matrix, $\widetilde{H}=i\sigma_2 H^*$ ($H$ being the SM Higgs doublet) and $L$ is the SM lepton doublet. 
Moreover, $M_N$  is a LN conserving matrix, while the mass matrix $\mu$ breaks LN explicitly by 2 units.

The right-handed neutrinos can be integrated out, 
leading to the Weinberg operator~\cite{Weinberg:1979sa} which generates masses for the light, active neutrinos of the form: 
\begin{equation}
  m_\nu\sim v^2 Y_\nu M_N^{-1}\mu (M_N^T)^{-1} Y^T_\nu. 
\end{equation}
Having TeV-scale right-handed neutrinos (e.g. motivated by naturalness~\cite{Casas:2004gh,Vissani:1997ys}) and $\mathcal{O}(1)$ Yukawa couplings would require $\mu\sim\mathcal{O}(\keV)$. In the original ISS formulation~\cite{Mohapatra:1986bd}, the smallness of this LNV parameter arises from a superstring inspired E6 scenario. Alternative explanations call upon other extensions of the SM such as Supersymmetry and Grand Unified Theories (see for instance~\cite{Malinsky:2005bi,Bazzocchi:2009kc}).  Here a dynamical origin for $\mu$ will be instead explored.
The $\mu$ parameter is technically natural: since it is the only parameter that breaks LN, its running is multiplicative and thus 
once chosen to be small, it will remain small at all energy scales.

To promote  the LN breaking parameter $\mu$ in the ISS scenario to a dynamical quantity, we choose to gauge the $B-L$ number~\cite{Mohapatra:1980qe}.  The spontaneous breaking of this symmetry will convey LN breaking, generate neutrino masses via a scalar vev, and give rise to a massive vector boson, dubbed here $Z'$. $B-L$ is an accidental symmetry of the SM, and it is well motivated in theories in which quarks and leptons are unified~\cite{Marshak:1979fm,Pati:1974yy,Georgi:1974my,Fritzsch:1974nn}. In unified theories, the chiral anomalies cancel within each family, provided that SM fermion singlets with charge $+1$ are included. In the usual ISS framework, this is not the case due to the presence of right-handed neutrinos with charges $+1$ and $-1$. The triangle anomalies that do not cancel are those involving three $U(1)_{B-L}$ vertices, as well as one $U(1)_{B-L}$ vertex and gravity. Therefore, to achieve anomaly cancellation for gauged $B-L$ we have to include additional chiral content to the model with charges that satisfy
\begin{align}
&\sum Q_i=0\Rightarrow\sum Q_{iL}-\sum Q_{iR}=0,\\
&\sum Q_i^3=0\Rightarrow\sum Q_{iL}^3-\sum Q_{iR}^3=0,
\end{align}
where the first and second equation refer to the mixed gravity-$U(1)_{B-L}$ and $U(1)_{B-L}^3$
 anomalies, respectively. The index $i$ runs through all fermions of the model.
 
In the following subsections we will discuss the fermion and the scalar sectors of the model in more detail.

\subsection{The fermion sector} 

Besides the anomaly constraint, the ISS mechanism can only work with a certain number of $N_R$ and $N_R'$ fields (see, e.g., Ref.~\cite{Abada:2014vea}). 
We find a phenomenologically interesting and viable scenario which consists of the following copies of SM fermion singlets and their respective $B-L$ charges: 
3 $N_R$ with charge $-1$; 3 $N_R'$ with charge $+1$; 1 $\chi_R$ with charge $+5$; 1 $\chi_L$ with charge $+4$ 
and 1 $\omega$ with charge $+4$\footnote{Introducing 2 $N_R$ and 3 $N_R'$ as for example in \cite{Abada:2014zra} leads to a keV sterile neutrino as a potentially interesting 
warm DM candidate~\cite{Adhikari:2016bei} in the spectrum due to the mismatch between the number of $N_R$ and $N_R'$. However, the 
relic abundance of this sterile neutrino, if thermally produced via freeze out, is an order of magnitude too large. Thus, in order to avoid 
its thermalisation, very small Yukawa couplings and mixings must be adopted instead. 
} Some of these right-handed neutrinos allow for a mass term, 
namely, $M_N \overline{N_R^c} N_R'$, but to lift the mass of the other sterile fermions and to generate SM neutrino masses, two extra scalars are introduced. 
Thus, besides the Higgs doublet $H$, the scalar fields $\phi_1$ with $B-L$ charge $+1$ and $\phi_2$ 
with charge $+2$ are considered. The SM leptons have $B-L$ charge $-1$, while the quarks have charge $1/3$. The 
scalar and fermion content of the model, related to neutrino mass generation, is summarised in 
Table~\ref{tab:particles}. The most general Lagrangian in the neutrino sector is then given by\footnote{Notice that a 
coupling $\phi_1^* {\overline{\omega}} Y_\omega \chi_R$, while allowed, can always be reabsorbed 
into $\phi_1^*  {\overline{\chi_L}} Y_\chi \chi_R$ through a rotation between $\omega$ and $\chi_L$.}
\begin{align}
- \mathcal{L}_\nu &=  \bar L Y_\nu \widetilde{H}  N_R +  {\overline{N_R^c}} M_N N_R' +  \phi_2  \overline{N_R^c} Y_N N_R + \phi_2^* \overline{(N_R')^c}\, Y'_N N_R'  +\phi_1^*  {\overline{\chi_L}} \, Y_\chi \chi_R +  {\rm h.c.}, 
\end{align}
 where the capitalised variables are to be understood as matrices (the indices were omitted). 

\begin{table}
\centering
\begin{tabular}{| c| c| c| c| c| c| c| c| c|}
\hline
  Particle & $ \phi_1$ & $ \phi_2$ & $ \nu_L $& $ N_R$ & $ N'_R$ & $ \chi_R$ & $\chi_L$& $\omega$\\ 
  \hline
  $U(1)_{B-L}$ charge & $ +1$ & $+2$ &$-1 $& $-1$ & $+1$ & $+5$ & $+4$& $+4$ \\
  \hline
  Multiplicity & $1$ & $1$ &$ 3 $& $3$ & $3$ & $1$ & $1$ & $1$\\
  \hline
\end{tabular}
\caption{Neutral fermions and singlet scalars with their $U(1)_{B-L}$ charge and their multiplicity. 
$\phi_{1,2}$ are SM singlet scalars while $N_R$, $N'_R$ and $\chi_R$ are right-handed and $ \chi_L$ and $\omega$ are left-handed SM singlet fermions respectively.}
\label{tab:particles}
\end{table}

The singlet fermion spectrum splits into two parts, an ISS sector composed by $\nu_L$, $N_R$, and $N'_R$, and a dark sector  with $\chi_L$ and $\chi_R$, as can be seen 
in the following mass matrix written in the basis $(\nu_L^c, N_R,N_R',\chi^c_L,\chi_R)$:
\begin{equation}
  M=\left(
  \begin{array}{c c c| c c}
  0&Y_\nu\widetilde{H}&0&0&0\\
  Y_\nu^T\widetilde{H}^\dagger&Y_N\phi_2&M_N&0&0\\
  0&M_N^T&Y_N'\phi_2^*&0&0\\
  \hline
  0&0&0&0& Y_\chi \phi_1^*\\
  0&0&0&Y_\chi^T\phi_1&0
  \end{array}\right).
\end{equation}
The dynamical equivalent of the $\mu$ parameter can be identified with $Y_N' \phi_2^*$\footnote{The analogous term $Y_N\phi_2$ - also dynamically generated - 
contributes to neutrino masses only at the one-loop level and is therefore typically sub-leading.}. 
After $\phi_1$ develops a vacuum expectation value (vev) a Dirac fermion $\chi=(\chi_L,\chi_R)$ and a massless fermion $\omega$ are formed in the dark sector.  
Although the cosmological impact of this extra relativistic degree of freedom may seem worrisome at first, we will show later that the contribution to 
$N_{\rm eff}$ is suppressed as this sector is well secluded from the SM.

To recover a TeV-scale ISS scenario with the correct neutrino masses and $\mathcal{O}(1)$ Yukawa couplings, 
$v_2\equiv\vev{\phi_2}\sim \keV\ll v$ (where $v=\vev{H}=246~\GeV$ is the electroweak vev) and $M_R\sim\TeV$ are needed. Moreover, the mass of the 
$B-L$ gauge boson will be linked to the vevs of $\phi_{1}$ and $\phi_{2}$, and hence to lift its mass above the electroweak scale will 
require $v_1\equiv\vev{\phi_1}\gtrsim~\TeV$. In particular, we will show that a triple scalar coupling $\eta\phi_1^2\phi_2^*$ can induce a small 
$v_2$ even when $v_1$ is large, similar to what occurs in the type-II seesaw~\cite{Magg:1980ut, Lazarides:1980nt, Mohapatra:1980yp, Schechter:1980gr, Cheng:1980qt}. 
After the spontaneous symmetry breaking, the particle spectrum would then consist of a $B-L$ gauge boson, 3 pseudo-Dirac neutrino pairs and a Dirac dark fermion at the TeV scale, as well as a massless dark fermion. The SM neutrinos would in turn develop small masses via the ISS in the usual way. Interestingly, both dark fermions only interact with the SM via the new gauge boson $Z'$ and via the suppressed mixing of $\phi_1$ with the Higgs.  They are also stable and thus the heavy dark fermion is a natural WIMP DM candidate.
Since all new fermions carry $B-L$ charge, they all couple to the $Z'$, but specially the ones in the dark sector which have larger $B-L$ charge.

\subsection{The scalar sector}
The scalar potential of the model can be written as
\begin{align}
  V&=\frac{m_H^2}{2} H^\dagger H+\frac{\lambda_H}{2} (H^\dagger H)^2 + \frac{m_1^2}{2} \phi_1^*\phi_1 + \frac{m_2^2}{2} \phi_2^*\phi_2
      + \frac{\lambda_1}{2}(\phi_1^*\phi_1)^2 + \frac{\lambda_2}{2}(\phi_2^*\phi_2)^2 \\
      &\quad+ \frac{\lambda_{12}}{2}(\phi_1^*\phi_1)(\phi_2^*\phi_2)+\frac{\lambda_{1H}}{2}(\phi_1^*\phi_1)(H^\dagger H)+\frac{\lambda_{2H}}{2}(\phi_2^*\phi_2)(H^\dagger H)-\eta(\phi_1^2\phi_2^*+\phi_1^{*2}\phi_2).\nonumber
\end{align}
Both $m_H^2$ and $m_1^2$ are negative, but $m_2^2$ is positive and large. Then, for suitable values of the quartic couplings, the vev of $\phi_2$, $v_2$, is only induced by the vev of $\phi_1$, $v_1$, through $\eta$ and thus it can be made small. 
With the convention $\phi_j=(v_j+\varphi_j+i\, a_j)/\sqrt{2}$ and the neutral component of the complex Higgs field given by $H^0=(v+h+i G_Z)/\sqrt{2}$ (where $G_Z$ is the Goldstone associated with the $Z$ boson mass),
the minimisation of the potential yields
\begin{align}
m_H^2 &= -\frac{1}{2}\left(\lambda_{1H}v_1^2+\lambda_{2H}v_2^2+2\lambda_H v^2\right)\simeq-\frac{1}{2}\left(\lambda_{1H}v_1^2+2\lambda_H v^2\right),\\
m_1^2 &= -\frac{1}{2}\left(2\lambda_1 v_1^2 + \lambda_{1H}v^2-4\sqrt{2}\eta v_2+\lambda_{12}v_2^2\right)\simeq-\frac{1}{2}\left(2\lambda_1 v_1^2 + \lambda_{1H}v^2\right),\\
m_2^2 &= \left(\frac{\sqrt{2}\eta}{v_2}-\frac{\lambda_{12}}{2}\right)v_1^2 -\lambda_2 v_2^2 - \frac{\lambda_{2H}}{2}v^2\simeq
		\frac{\sqrt{2}\eta v_1^2}{v_2},
\end{align}
or, equivalently,
\begin{equation}
  v_2\simeq \frac{\sqrt{2}\eta v_1^2}{m_2^2}~.
\end{equation}
Clearly, when $\eta\to0$ or $m_2^2\to\infty$, the vev of $\phi_2$ goes to zero. For example, to obtain $v_2\sim\mathcal{O}(\keV)$, one could have $m_2\sim 10~ {\rm TeV}$, $v_1\sim 10~ \TeV$, 
and $\eta\sim 10^{-5}~\GeV$.
The neutral scalar mass matrix is then given by
\begin{equation}
  M_0^2\simeq\left(
  \begin{array}{c c c}
  \lambda_H v^2 & \lambda_{1H}v_1 v/2&0\\
  \lambda_{1H}v_1 v/2&\lambda_1 v_1^2&-\sqrt{2}\eta v_1\\
  0&-\sqrt{2}\eta v_1& \eta v_1^2/\sqrt{2} v_2
  \end{array}\right).
\end{equation}
Higgs data constrain the mixing angle between ${\rm Re}(H^0)$ and ${\rm Re}(\phi_1^0)$ to be below $\sim30\%$~\cite{Robens:2016xkb}. Moreover, since $\eta\ll m_2,v_1$, the mixing between the new scalars is also small. Thus, the masses of the physical scalars $h$, $\varphi_1$ and $\varphi_2$ are approximately 
\begin{equation}
  m_h^2=\lambda_H v^2,\quad m_{\varphi_1}^2=\lambda_1 v_1^2,\quad{\rm and}\quad m_{\varphi_2}^2=m_2^2/2,
\end{equation}
while the mixing angles $\alpha_1$ and $\alpha_2$ between $h-\varphi_1$ and $\varphi_1-\varphi_2$, respectively, are 
\begin{equation}
  \tan\alpha_1\simeq \frac{\lambda_{1H}}{\lambda_1}\frac{v}{2v_1},\quad{\rm and}\quad
  \tan\alpha_2\simeq2\frac{v_2}{v_1}.
  \label{eq:tanalpha}
\end{equation}
If $v_1\sim\TeV$ and the quartics $\lambda_1$ and $\lambda_{1H}$ are $\mathcal{O}(1)$, the mixing $\alpha_1$ is expected to be small but non-negligible. A mixing between the Higgs doublet and a scalar singlet can only diminish the Higgs couplings to SM particles. Concretely, the couplings of the Higgs to gauge bosons and fermions, relative to the SM couplings, are
\begin{equation}
  \kappa_F=\kappa_V=\cos\alpha_1,
\end{equation}
which is constrained to be  $\cos{\alpha_1}>0.92$ (or equivalently $\sin{\alpha_1}<0.39$)~\cite{Khachatryan:2016vau}.
Since the massless fermion does not couple to any scalar, and all other extra particles in the model are heavy, the modifications to the SM Higgs couplings are the only phenomenological impact of the model on Higgs physics. The other mixing angle, $\alpha_2$, is very small since it is proportional to the LN breaking vev and thus is related to neutrino masses. Its presence will induce a mixing between the Higgs and $\varphi_2$, but for the parameters of interest here it is unobservable.\\
 Besides Higgs physics, the direct production of $\varphi_1$ at LHC via its mixing with the Higgs would be possible if it is light enough. Otherwise, loop effects that would change the $W$ mass bound can also test this scenario imposing $\sin\alpha_1\lesssim 0.2$ for $m_{\varphi_1}=800~$GeV~\cite{Robens:2016xkb}. 

Apart from that, the only physical pseudoscalar degree of freedom is 
\begin{equation}
  A=\frac{1}{\sqrt{v_1^2 + 4 v_2^2}}\left[2 v_2 a_1-v_1 a_2\right]
\end{equation}
and its mass is degenerate with the heavy scalar mass, $m_A\simeq m_{\varphi_2}$.\\

We have built this model in {\tt SARAH} 4.9~\cite{Staub:2012pb,Staub:2013tta,Staub:2015kfa,Vicente:2015zba}. 
This Mathematica package produces the model files for {\tt SPheno} 3.3.8~\cite{Porod:2003um,Porod:2011nf} and {\tt CalcHep}~\cite{Belyaev:2012qa} 
which are then used to study the DM phenomenology with {\tt Micromegas} 4.3~\cite{Belanger:2014vza}. 
We have used these packages to compute the results presented in the following sections. Moreover, we will present analytical estimations to further interpret the numerical results.

\section{Dark matter phenomenology}
\label{sec:DM}
As discussed in the previous section, in this dynamical realisation of the ISS mechanism we have two stable fermions. One of them is a Dirac fermion, $\chi=(\chi_L,\chi_R)$, which 
acquires a mass from $\phi_1$, and therefore is manifest at the TeV scale. The other, $\omega$, is massless and will contribute to the number of relativistic species in the early Universe. First we analyse if $\chi$ can yield 
the observed DM abundance of the Universe.\\

\subsection{Relic density}
In the early Universe, $\chi$ is  in thermal equilibrium with the plasma due to its gauge interaction with $Z'$. The relevant part of the Lagrangian is

\begin{equation}
  \mathcal{L}_{DM} = - g_{\rm BL} \bar\chi \gamma^\mu ( 5P_R+4 P_L) \chi Z'_\mu + \frac{1}{2}M_{Z'}^2Z'_\mu Z^{\prime \mu} - m_\chi \bar\chi\chi,
\end{equation}
where
\begin{equation}
 M_{Z'}= g_{\rm BL} \sqrt{v_1^2+4v_2^2}\simeq g_{\rm BL} v_1,~~{\rm and}~~m_\chi = Y_\chi v_1/\sqrt{2},
\end{equation}
 
and $P_{R,L}$ are the chirality projectors.

 \begin{figure}
\centering
\includegraphics[scale=0.4]{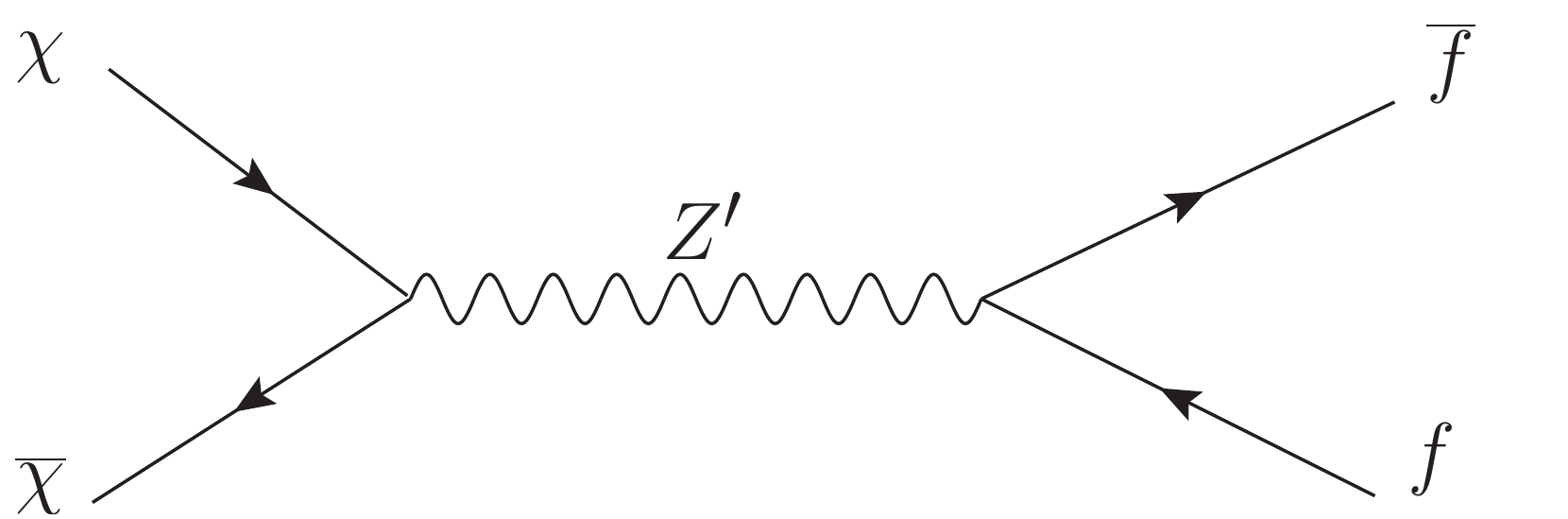}\vspace{1mm}
\includegraphics[scale=0.4]{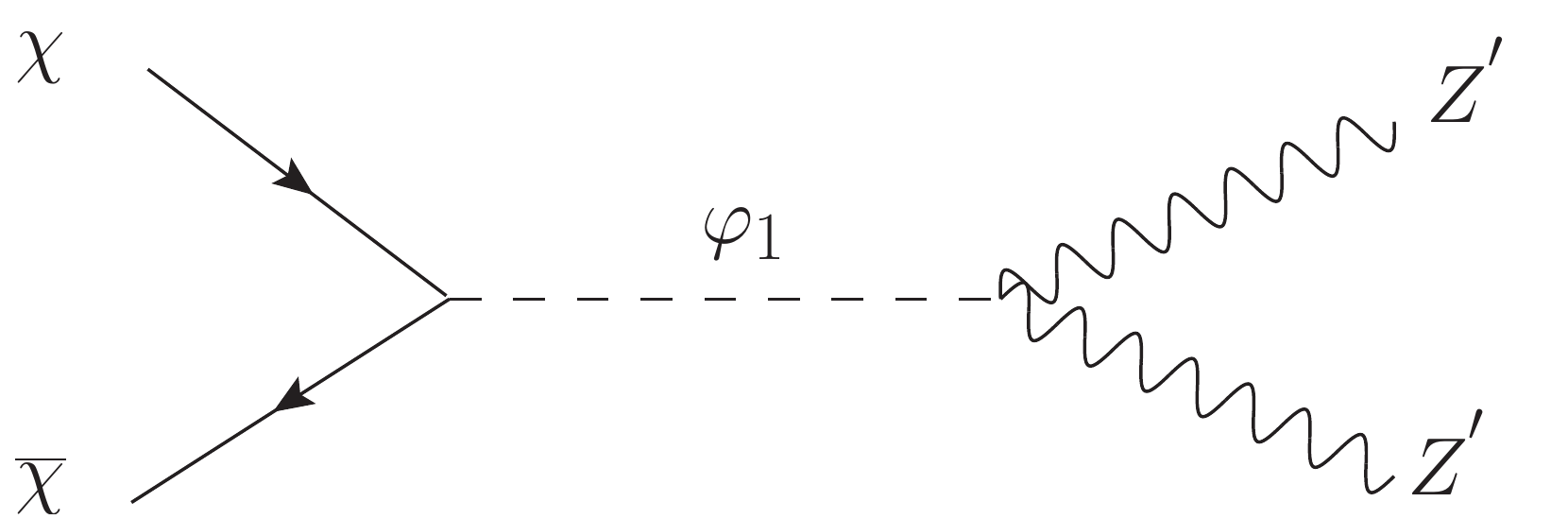}\vspace{2mm}
\includegraphics[scale=0.4]{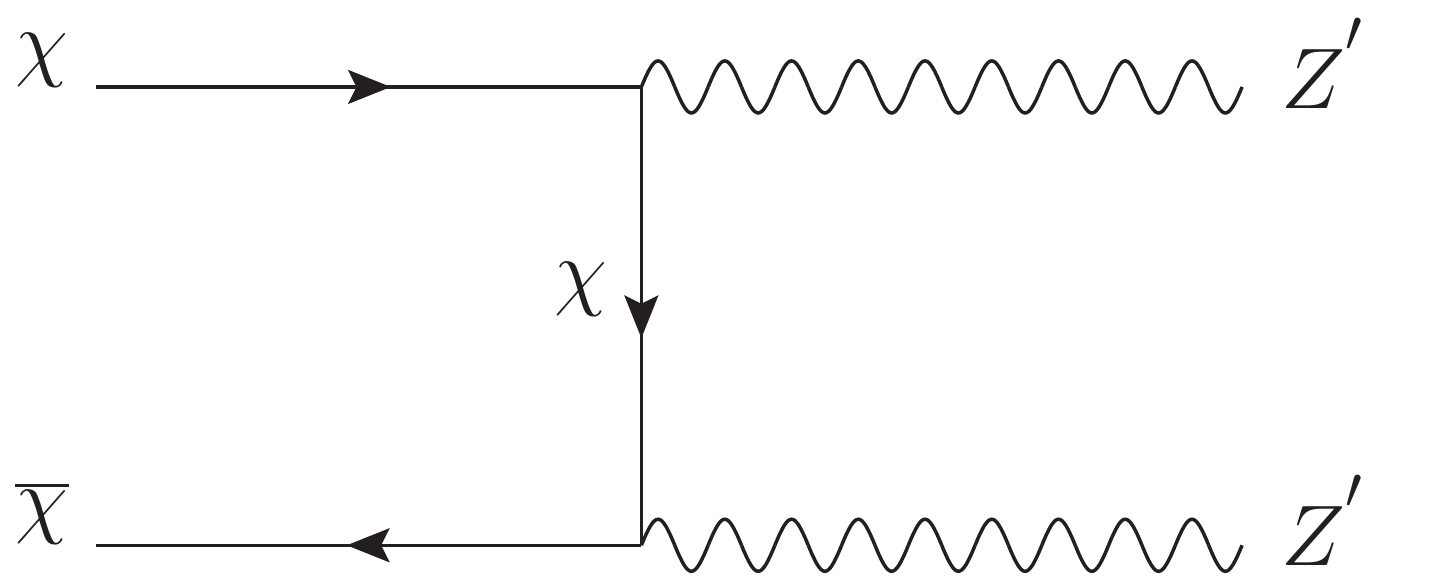}
\includegraphics[scale=0.4]{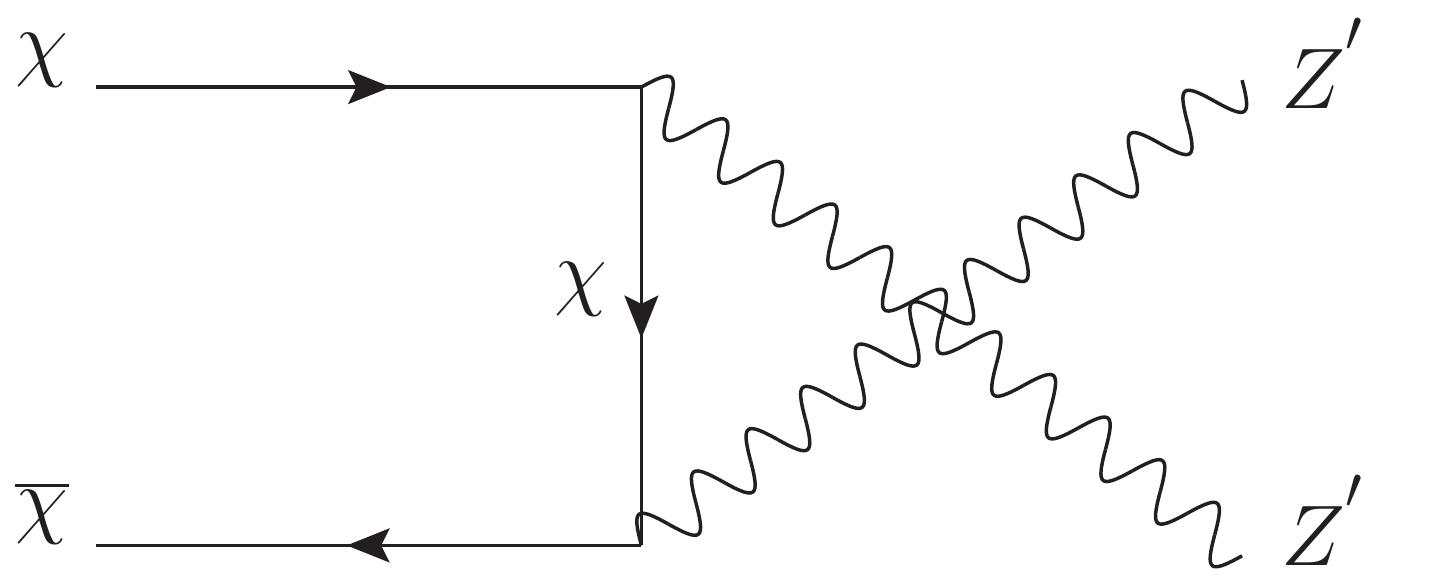}
\caption{\label{fig:DManni} DM annihilation channels $\chi\bar{\chi}\to f  \bar f$ via the $Z'$ boson and $\chi\bar{\chi}\to Z' Z'$. The $\chi\bar{\chi}\to Z' Z'$ channel opens up when  $M_{Z'}^2 < m_\chi^2$.   Since the process $\chi\bar{\chi}\to \varphi_1 \to Z' Z'$ is velocity suppressed this diagram is typically  subleading.}
\end{figure}

The main annihilation channels of $\chi$ are $\chi \bar{\chi}\to f \bar f$ via the $Z'$ boson exchange 
and $\chi\bar{\chi}\to Z' Z'$ - if kinematically allowed (see fig.~\ref{fig:DManni}).
 
The annihilation cross section to a fermion species $f$, at leading order in $v$, reads:
\begin{equation}
  \vev{\sigma \mathrm{v}}_{ff} \simeq n_c (q_{\chi_{L}}+q_{\chi_{R}})^2~\frac{q^2_{f_L}+q^2_{f_R}}{8\pi}\frac{g_{\rm BL}^4 m_\chi^2}{(4m_\chi^2-M_{Z'}^2)^2+\Gamma^2_{Z'}M_{Z'}^2} + 
  \mathcal{O}\left(v^2 \right),
	\label{eq:fermions}
\end{equation}
see e.g.~\cite{Alves:2015mua,Lindner:2010rr}, 
where $n_c$ is the color factor of the final state fermion (=1 for leptons), $q_{\chi_{L}}=4$ and $q_{\chi_{R}}=5$ and $q_{f_{L,R}}$ are the $B-L$ charges of the left- and 
right-handed components of the DM candidate $\chi$ and of the fermion $f$, respectively. Moreover, the partial decay 
width of the $Z'$ into a pair of fermions (including the DM, for which $f=\chi$) is given by 

\begin{equation}
 \Gamma_{Z'}^{ff} =  n_c~ g_{\rm BL}^2 \frac{\left(6 q_{f_L} q_{f_R} m^2_f + \left( q^2_{f_L}+q^2_{f_R} \right) 
 \left(M_{Z'}^2 - m_f^2 \right) \right) \sqrt{M^2_{Z'} - 4 m_f^2}}{24 \pi M^2_{Z'}}\,.
	\label{eq:width}
\end{equation}

When $M_{Z'}^2 < m_\chi^2$, the annihilation channel $\chi\bar{\chi}\to Z'Z'$ is also available. The cross section for this process (lower diagrams in fig.~\ref{fig:DManni}) 
is given by  (to leading order in the relative velocity) \cite{Alves:2015mua}
\begin{align}
  \vev{\sigma \mathrm{v}}_{Z'Z'}&\simeq\frac{1}{256\pi m_\chi^2 M_{Z'}^2}
  \left(1-\frac{M_{Z'}^2}{m_\chi^2}\right)^{3/2}\left(1-\frac{M_{Z'}^2}{2 m_\chi^2}\right)^{-2} \nonumber\\
  &\left(8 g_{\rm{BL}}^4 (q_{\chi_{R}}+q_{\chi_{L}})^2 (q_{\chi_{R}}-q_{\chi_{L}})^2 m_\chi^2+\left( (q_{\chi_{R}}-q_{\chi_{L}})^4+  (q_{\chi_{R}}+q_{\chi_{L}})^4 \right. \right. \nonumber \\
  &\left.\left.-6  (q_{\chi_{R}}-q_{\chi_{L}})^2  (q_{\chi_{R}}+q_{\chi_{L}})^2\right) g_{\rm{BL}}^4 M_{Z'}^2 \right)~,
		\label{eq:Zs}
  \end{align}

The $\chi\bar{\chi}\to\varphi_1\to Z{'} Z{'} $ (upper right diagram in fig.~\ref{fig:DManni}) channel is velocity suppressed and hence typically subleading. 
Further decay channels like  $\chi\bar{\chi}\to \varphi_1 \varphi_1$ and $\chi\bar{\chi}\to Z' \varphi_1$ open when $2 m_\chi>m_{\varphi_1}+m_{\varphi_1} (m_{\varphi_1}+m_{Z^{'}})$. 
With $m_\chi= Y_\chi/\sqrt{2}v_1,$ $m_{\varphi_1}=\sqrt{\lambda_1}v_1,$ $m_{Z^{'}}=g_{\rm BL}v_1$ and the additional constraint from perturbativity  $Y_\chi\leq 1$ we get only 
small kinematically allowed   regions which play a subleading role for the relic abundance. 
The cross section for the annihilation channel $\chi\bar{\chi}\to Z' h^0$ is also subleading due to the mixing angle $\alpha_1$ between $\varphi_1 - h^0$  which is small although 
non-negligible (cf. Eq.~\eqref{eq:tanalpha}).

The relic density of $\chi$ has been computed numerically with {\tt Micromegas} obtaining also, for several points of the parameter space, the DM freeze-out temperature at which the annihilation rate becomes smaller than the Hubble rate $\vev{\sigma \mathrm{v}} n_{\chi} \lesssim H$. Given the freeze-out temperature and the annihilation cross sections of Eqs.~\eqref{eq:fermions} and~\eqref{eq:Zs}, the DM relic density can thus be estimated by~\cite{Kolb:1990vq}:

\begin{equation}
  \Omega_\chi h^2 = \frac{2.5\cdot 10^{28} m_\chi}{T^{\rm f.o.}_\chi M^2_{Pl}\sqrt{g_\star}\vev{\sigma \mathrm{v}}},
\end{equation}
where $g_\star$ is the number of degrees of freedom in radiation at the temperature of freeze-out of the DM 
($T^{\rm f.o.}_\chi$), $\vev{\sigma \mathrm{v}}$ is its thermally averaged annihilation cross section and $M_{Pl} = 1.2 \cdot 10^{19}$ GeV is the Planck mass. In Sec.~\ref{sec:results} we will use this estimation of $\Omega_\chi h^2$ together with its constraint $\Omega_\chi h^2 \simeq 0.1186 \pm 0.0020$~\cite{Ade:2015xua,Olive:2016xmw} to explore the regions of the parameter space for which the correct DM relic abundance is obtained.

\subsection{Direct Detection}
The same $Z'$ couplings that contribute to the relic abundance can give rise to signals in DM direct detection experiments.
The DM-SM interactions in the model via the $Z'$ are either vector-vector or axial-vector interactions. 
Indeed, the $Z'$- SM interactions are vectorial (with the exception of the couplings to neutrinos) while $\chi$ has different left- and right-handed charges. 
The axial-vector interaction does not lead to a signal in direct detection and the vector-vector 
interaction leads to a spin-independent cross section \cite{Cheung:2012gi}. 

The  cross section for coherent elastic scattering on a nucleon is
\begin{align}
\sigma^{\rm DD}_{\chi}=\frac{\mu_{\chi \rm N}^2}{\pi}\left(\frac{9}{2}\frac{g_{\rm BL}^2}{M_{Z'}^2}\right)^2
\label{sigdd}
\end{align}
 where $\mu_{\chi \rm N}$ is the reduced mass of the DM-nucleon system.
The strongest bounds on the spin-independent scattering cross section come from LUX~\cite{Akerib:2016vxi} and XENON1T~\cite{Aprile:2017iyp}. The constraint on the DM-nucleon  
scattering cross section is  $\sigma^{\rm DD}_{\chi}<10^{-9}$~pb for $m_{\chi}=1$~TeV  and $\sigma^{\rm DD}_{\chi}<10^{-8}$~pb  for $m_\chi=10$~TeV. 
The experimental bound on the spin-independent cross section (Eq.~\eqref{sigdd}) allows to derive a lower bound on the vev of $\phi_1$:

\begin{equation}
v_1 ~\text{[GeV]}> \left(\frac{2.2\cdot 10^9}{\sigma^{\rm DD}_{\chi}~\text{[pb]}} \right)^{1/4}~.
\end{equation}

This bound pushes the DM mass to be $m_\chi \gtrsim$ TeV. 
For instance, for $g_{\rm BL} = 0.25$ and $m_{Z'} = 10$ TeV, a DM mass $m_\chi = 3.8$ TeV is required to have 
  $\sigma^{\rm DD}_{\chi} ~\sim 9 \times 10^{-10}$ pb. In turn, this bound translates into a lower limit on the vev of $\phi_1$: $v_1 \gtrsim 40$ TeV (with $Y_\chi \gtrsim 0.1$). 
Next generation experiments such as XENON1T~\cite{Aprile:2015uzo} and LZ~\cite{Akerib:2015cja} are expected to improve the current bounds by an order of magnitude and could test the parameter space of this model, as it will be discussed in Sec.~\ref{sec:results}. 

\subsection{Indirect Detection}
In full generality, the annihilation of  $\chi$ today could lead also to indirect detection signatures, in the form of charged cosmic rays, neutrinos and  gamma rays. 
However, since the main annihilation channel of $\chi$ is via the $Z'$ which couples dominantly to the dark sector, the bounds from indirect detection searches turn out to be subdominant.

The strongest experimental bounds come from gamma rays produced through direct emission from the annihilation of $\chi$ into $\tau^+ \tau^-$. Both the constraints from the Fermi-LAT Space Telescope (6-year observation of gamma rays from dwarf spheroidal galaxies)~\cite{Ackermann:2015zua} and H.E.S.S. (10-year observation of gamma rays from the Galactic Center)~\cite{Abramowski:2011hc} are not very stringent for the range of DM masses considered here. 
Indeed, the current experimental bounds on the velocity-weighted annihilation cross section $<\sigma v> (\chi \bar{\chi}\to \tau^+\tau^-)$ range from $10^{-25}~\text{cm}^3 \text{s}^{-1}$ to $10^{-22}~\text{cm}^3 \text{s}^{-1}$ for DM masses between 1 and 10~TeV. These values are more than two orders of magnitude above the values obtained for the regions of the parameter space in which we obtain the correct relic abundance (notice that the branching ratio of the DM annihilation to $\chi$ into $\tau^+ \tau^-$ is only about $5\%$).
Future experiments like CTA~\cite{Wood:2013taa} could be suited to sensitively address DM masses in the range of interest of this model ($m_\chi \gtrsim 1$ TeV).\\

\subsection{Effective number of neutrino species, $\boldmath{N_{\rm eff}}$}
The presence of the massless fermion $\omega$ implies a contribution to the number of relativistic degrees of freedom in the early Universe.
In the following, we discuss its contribution to the effective number of neutrino species, $N_{\rm eff}$, which has been measured to be $N_{\rm eff}^{exp}=3.04\pm 0.33$ \cite{Ade:2015xua}. 
Since the massless $\omega$ only interacts with the SM via the $Z'$, its contribution to $N_{\rm eff}$ will be washed out through entropy injection to the thermal bath by the number of relativistic degrees of freedom $g_\star(T)$ at the time of its decoupling:  
\begin{align}
\Delta N_{\rm eff}=\left(\frac{T^{\rm f.o.}_\omega}{T_\nu}\right)^4 ~= \left(\frac{11}{2 g_\star(T^{\rm f.o.}_\omega)}\right)^{4/3}~,
\end{align}
where $T^{\rm f.o.}_\omega$ is the freeze-out temperature of $\omega$ and $T_\nu$ is the temperature of the neutrino background. 
The freeze-out temperature can be estimated when the Hubble expansion rate of the Universe $H = 1.66 \sqrt{g_\star} T^2/M_{Pl}$ overcomes the $\omega$ interaction rate $\Gamma = <\sigma v> n_\omega$ leading to:
\begin{align}
(T^{\rm f.o.}_\omega)^3 \sim \frac{2.16 \sqrt{g_\star}M^4_{Z'}}{M_{Pl} g_{\rm BL}^4 \sum_f (q^2_{f_L} + q^2_{f_R})}~.
\end{align}

With the typical values that satisfy the correct DM relic abundance: $m_{Z'}\sim\mathcal{O}(10$ TeV) and $g_{\rm BL}\sim\mathcal{O}$(0.1) $\omega$ would therefore freeze out at $T^{\rm f.o.}_\omega \sim 4$~GeV, before the QCD phase transition. Thus, the SM bath will heat significantly after $\omega$ decouples and the contribution of the latter to the number of degrees of freedom in radiation will be suppressed:
\begin{align}
\Delta N_{\rm eff} \approx 0.026
\end{align}
which is one order of magnitude smaller than the current uncertainty on $N_{\rm eff}$. 
For gauge boson masses between 1-50 TeV and gauge couplings between 0.01 and 0.5, $\Delta N_{\rm eff}\in[0.02,0.04]$. 
Nevertheless, this deviation from $N_{\rm eff}$ matches the sensitivity expected from a EUCLID-like survey \cite{Basse:2013zua, Amendola:2016saw} and would be 
an interesting probe of the model in the future.

 \section{Collider phenomenology}
 \label{sec:colliders}
The new gauge boson can lead to resonant signals at the LHC. Dissimilarly from the widely studied case of a sequential $Z'$ boson, where the new boson decays dominantly to dijets, the elusive $Z'$ couples more strongly to leptons than to quarks (due to the $B-L$ number). Furthermore, it has large couplings to the SM singlets, specially $\chi$ and $\omega$ which carry large $B-L$ charges.
Thus, typical branching ratios are
$\sim$70\% invisible (i.e. into SM neutrinos and $\omega$), $\sim$12\% to quarks and $\sim$18\% to charged leptons.\footnote{If the decay 
channels to the other SM singlets are kinematically accessible, 
specially into $\chi$ and into the  $N_R, N'_R$ pseudo-Dirac pairs, the invisible branching ratio can go up to $\sim 87\%$, making the $Z'$ even more elusive and rendering these collider constraints irrelevant with respect to direct DM searches.} 
LHC  $Z'\to e^+e^-,\mu^+\mu^-$
resonant searches~\cite{Khachatryan:2016zqb, ATLAS:2016cyf} can be easily recast into constraints on the elusive $Z'$. The production cross section times branching ratio to dileptons 
is given by
\begin{equation}
  \sigma(pp\to Z'\to\ell \bar \ell)=\sum_q\frac{C_{qq}}{s M_{Z'}}\Gamma(Z'\to q \bar q){\rm BR}(Z'\to\ell \bar \ell),
\end{equation}
where $s$ is the center of mass energy, $\Gamma(Z'\to q \bar q)$ is the partial width to $q \bar q$ pair given by Eq.~\eqref{eq:width}, 
and $C_{qq}$ is the $q \bar q$ luminosity function obtained here using the parton distribution function MSTW2008NLO~\cite{Martin:2009iq}.
To have some insight on what to expect, we compare our $Z'$ with the usual sequential standard model (SSM) $Z'$, in which all couplings to fermions are equal to the $Z$ couplings.
The dominant production mode is again $q \bar q\to Z'$ though the coupling in our case is mostly vectorial.  The main dissimilarity arrives from the branching ratio to dileptons, as there are many additional fermions charged under the new gauge group. 
In summary, only $\mathcal{O}(1)$ differences in the gauge coupling bounds are expected, between the SSM $Z'$ and our elusive $Z'$.

\begin{figure}
\centering
\includegraphics[scale=0.5]{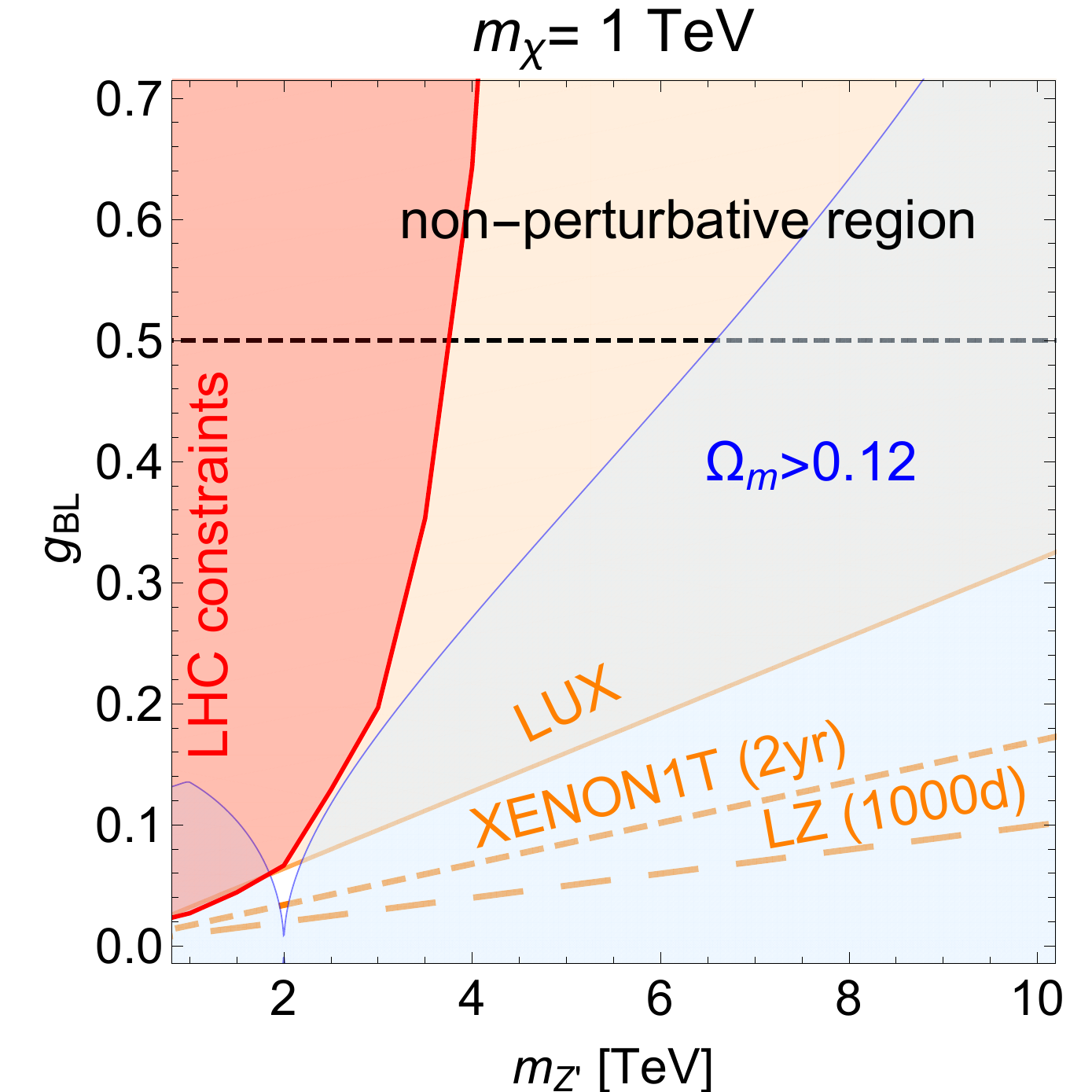}
\includegraphics[scale=0.5]{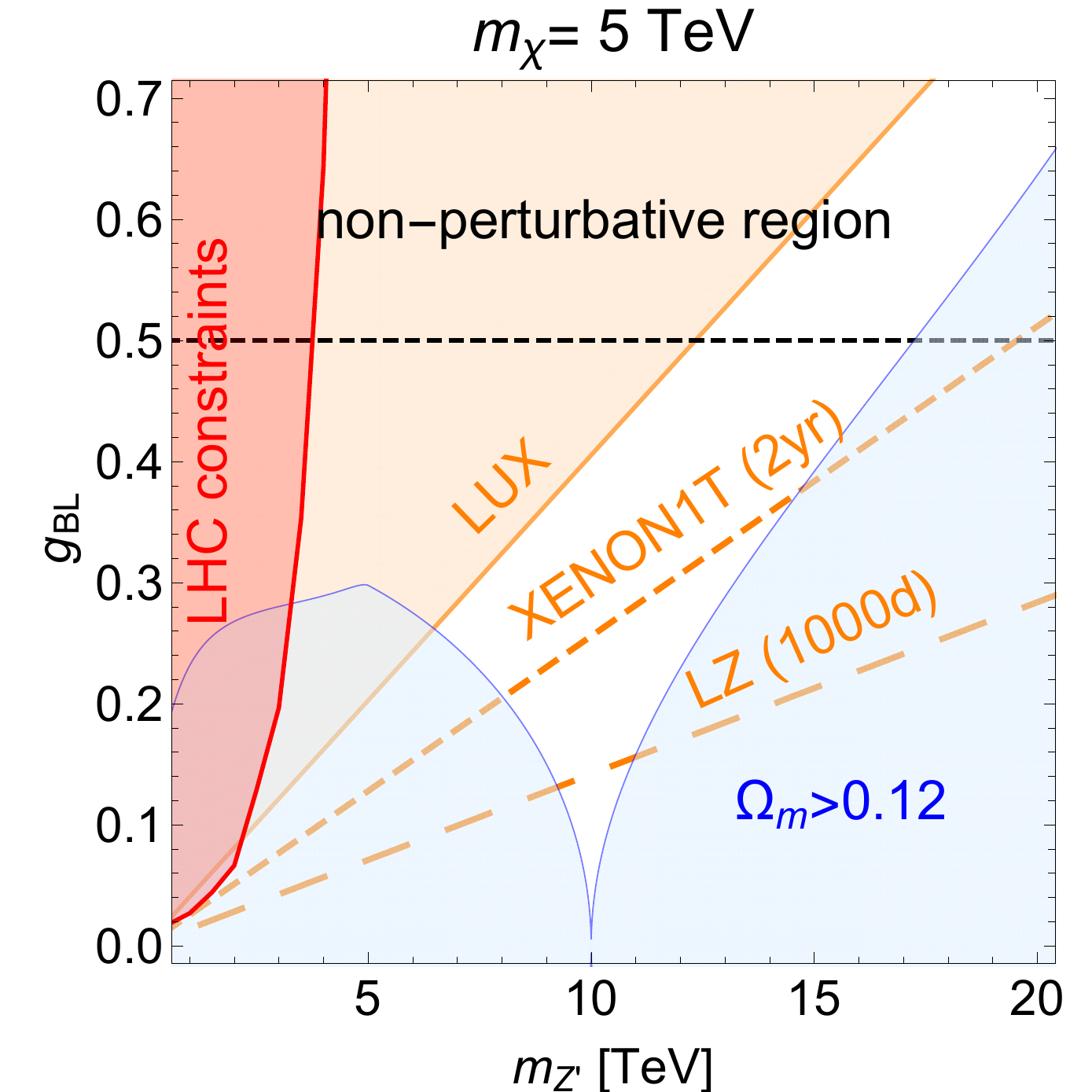}\vspace{2mm}
\includegraphics[scale=0.5]{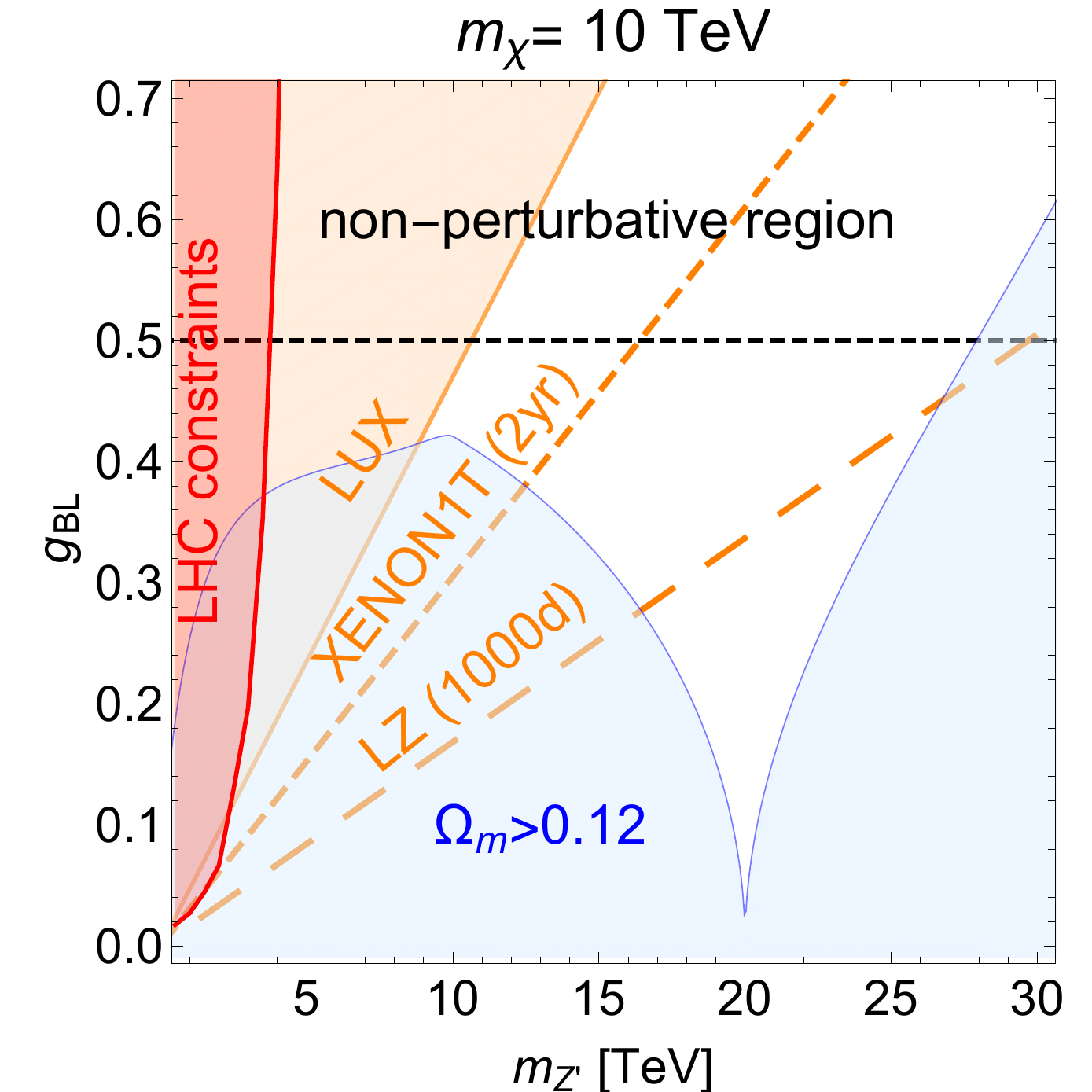}
\includegraphics[scale=0.5]{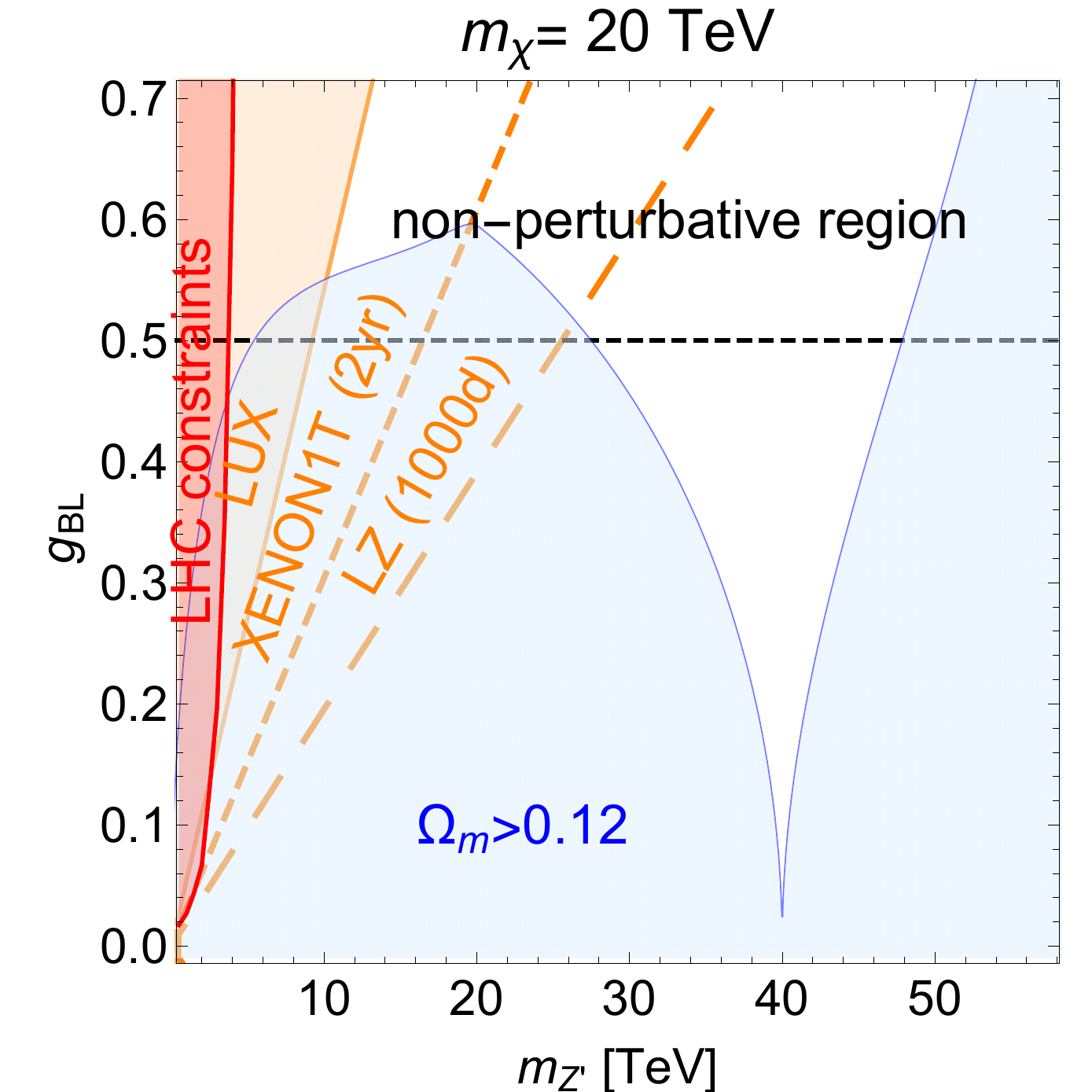}
\caption{\label{fig:g_vs_mz} Summary plots of our results.  The red region to the left is excluded by LHC constraints on the $Z'$ (see text for details), the region above $g_{\rm BL}>0.5$ is 
non-perturbative due to  $g_{\rm BL}\cdot q_{\rm max}\leq\sqrt{2\pi}$. In the blue shaded region DM is overabundant. The orange coloured region is already excluded by direct 
detection constraints from LUX~\cite{Akerib:2016vxi}, the short-dashed line indicates the future constraints from 
XENON1T~\cite{Aprile:2015uzo} (projected sensitivity assuming $2t \cdot y$), the long-dashed line the future constraints from LZ~\cite{Akerib:2015cja} 
(projected sensitivity for 1000d of data taking).}
\end{figure}

\vspace{0.5cm}
\section{Results}
\label{sec:results}
We now combine in fig.~\ref{fig:g_vs_mz} the constraints coming from DM relic abundance, DM direct detection experiments and collider searches. We can clearly see the synergy between these different observables. Since the DM candidate in our model is a thermal WIMP, the relic abundance constraint puts a lower bound on the gauge coupling, excluding the blue shaded region in the panels of fig.~\ref{fig:g_vs_mz}. On the other hand, LHC resonant searches essentially put a lower bound on the mass of the $Z'$ (red shaded region), while 
the LUX direct detection experiment constrains the product $g_{\rm BL} \cdot M_{Z'}$ from above (orange shaded region). For reference, we also show the prospects for future direct detection experiments, 
namely, XENON1T (orange short-dashed line, projected sensitivity assuming $2t \cdot y$) and LZ (orange long-dashed line, projected sensitivity for 1000d of data taking). Finally, if the gauge coupling is too large, perturbativity will be lost. 
To estimate this region we adopt the constraint $g_{\rm BL}\cdot q_{\rm max}\leq\sqrt{2\pi}$ and being the  largest $B-L$ charge $q_{\rm max}=5$, we obtain $g_{\rm BL}>0.5$ for the non-perturbative region. The white region in these panels represents the allowed region. We present four different DM masses so as to exemplify the dependence on $m_\chi$. First, we see that for DM masses at 1~TeV (upper left panel), there is only a tiny allowed region in which the relic abundance is set via resonant $\chi \bar{\chi}\to Z'\to f \bar f$ annihilation. For larger masses, the allowed region grows but some amount of enhancement is in any case needed so that the $Z'$ mass needs to be around twice the DM mass in order to obtain the correct relic abundance. For $m_\chi$ above 20~TeV (lower right panel), the allowed parameter space cannot be fully probed even with generation-2 DM direct detection experiments. 
 
On top of the DM and collider phenomenology discussed here, this model allows for a rich phenomenology in other sectors.
 In full analogy to the standard ISS model, the dynamical ISS mechanism here considered is also capable of generating 
 a large CP asymmetry in the lepton sector at the TeV scale, thus allowing for a possible  
 explanation of the baryon asymmetry of the Universe via {\textit{leptogenesis}}~\cite{Dev:2009aw,Blanchet:2010kw,Abada:2015rta,Hernandez:2015wna}. \\
 Moreover, the heavy sterile states typically introduced in ISS scenarios, namely the three pseudo-Dirac pairs from the states
 $N_R$ and $N_R^{'} $ can lead to new contributions to a wide array of observables~\cite{Shrock:1980vy,Schechter:1980gr,Shrock:1980ct,Shrock:1981wq,Langacker:1988ur,Bilenky:1992wv,Nardi:1994iv,Tommasini:1995ii,Antusch:2006vwa,Antusch:2008tz,Biggio:2008in,Forero:2011pc,Abdallah:2011ew,Alonso:2012ji,Boucenna:2014zba,Abada:2014nwa,Abada:2014cca,Arganda:2014dta,Abada:2015trh,Abada:2016awd,Abada:2015oba,Abada:2015zea,Fernandez-Martinez:2015hxa,DeRomeri:2016gum,Abada:2016vzu} such as weak universality, lepton flavour violating or precision electroweak observables, which allow to constrain the mixing of the SM neutrinos with the extra heavy pseudo-Dirac pairs to the level of $10^{-2}$ or even better for some elements~\cite{Antusch:2014woa,Fernandez-Martinez:2016lgt}. \\

\section{Conclusions}
\label{sec:conclusions}

The simplest extension to the SM particle content so as to accommodate the experimental evidence for neutrino masses and mixings is the addition of right-handed neutrinos, 
making the neutrino sector more symmetric to its charged lepton and quark counterparts. In this context, the popular Seesaw mechanism also gives a rationale 
for the extreme smallness of these neutrino masses as compared to the rest of the SM fermions 
through a hierarchy between two different energy scales: 
the electroweak scale -- at which Dirac neutrino masses are induced -- and a much larger energy scale tantalizingly close to the Grand Unification scale 
at which Lepton Number is explicitly broken by the Majorana mass of the right-handed neutrinos. 
On the other hand, this very natural option to explain the smallness of neutrino masses automatically makes the mass of the Higgs extremely unnatural, given the hierarchy problem that is hence introduced between the electroweak  scale and the heavy Seesaw scale. 

The ISS mechanism provides an elegant solution to this tension by lowering the Seesaw scale close to the electroweak  scale, thus avoiding the Higgs hierarchy problem altogether. In the ISS the smallness of neutrino masses is thus not explained by a strong hierarchy between these scales but rather by a symmetry argument. Since neutrino masses are protected by the Lepton Number symmetry, or rather $B-L$ in its non-anomalous version, if this symmetry is only mildly broken, neutrino masses will be naturally suppressed by the small parameters breaking this symmetry. In this work, the possibility of breaking this gauged symmetry dynamically has been explored. 

Since the ISS mechanism requires a chiral structure of the extra right-handed neutrinos under the $B-L$ symmetry, some extra states are predicted for this symmetry to be gauged due to anomaly cancellation. The minimal such extension requires the addition of three new fields with large non-trivial $B-L$ charges. Upon the spontaneous breaking of the $B-L$ symmetry, two of these extra fields become a massive heavy fermion around the TeV scale while the third remains massless. Given their large charges, the $Z'$ gauge boson mediating the $B-L$ symmetry couples preferentially to this new \textit{dark sector} and much more weakly to the SM leptons and particularly to quarks, making it rather \textit{elusive}. 

The phenomenology of this new dark sector and the elusive $Z'$ has been investigated. We find that the heavy Dirac fermion is a viable DM candidate in some regions of the parameter space. While the elusive nature of the heavy $Z'$ makes its search rather challenging at the LHC, it would also mediate spin-independent direct detection cross sections for the DM candidate, which place very stringent constraints in the scenario. Given its preference to couple to the dark sector and its suppressed couplings to quarks, the strong tension between direct detection searches and the correct relic abundance for $Z'$ mediated DM is mildly alleviated and some parts of the parameter space, not far from the resonance, survive present constraints. Future DM searches by XENON1T and LZ will be able to constrain this possibility even further. Finally, the massless dark fermion will contribute to the amount of relativistic degrees of freedom in the early Universe. While its contribution to the effective number of neutrinos is too small to be constrained with present data, future EUCLID-like surveys could reach a sensitivity close to their expected contribution, making this alternative probe a promising complementary way to test this scenario.

\section*{Acknowledgements}
VDR would like to thank A. Vicente for valuable assistance on SARAH and SPheno. JG would like to thank Fermilab for kind hospitality during the final stages of this project.
This work is supported in part by the EU grants H2020-MSCA-ITN-2015/674896-Elusives and H2020-MSCA-2015-690575-InvisiblesPlus. 
VDR acknowledges support by the Spanish grant SEV-2014-0398 (MINECO) and partial support by the Spanish grants FPA2014-58183-P, Multidark CSD2009-00064 and PROMETEOII/2014/084
(Generalitat Valenciana). EFM acknowledges support from the EU FP7 Marie Curie Actions CIG NeuProbes (PCIG11-GA-2012-321582), "Spanish Agencia Estatal de Investigaci\'on" (AEI) 
and the EU "Fondo Europeo de Desarrollo Regional" (FEDER) through the project FPA2016-78645-P and the Spanish MINECO through the ``Ram\'on y Cajal'' programme (RYC2011-07710) 
and through the Centro de Excelencia Severo Ochoa Program under grant SEV-2012-0249 and the HPC-Hydra cluster at IFT. 
The work of VN was supported by the SFB-Transregio TR33 ``The Dark Universe".
This manuscript has been authored by Fermi Research Alliance, LLC under Contract No. DE-AC02-07CH11359 with the U.S. Department of Energy, Office of Science, Office of High Energy Physics. The United States Government retains and the publisher, by accepting the article for publication, acknowledges that the United States Government retains a non-exclusive, paid-up, irrevocable, world-wide license to publish or reproduce the published form of this manuscript, or allow others to do so, for United States Government purposes.

\bibliographystyle{JHEP}
\bibliography{paper_IFT}

\end{document}